\documentstyle[aps,prl,epsf,preprint,cite,amsmath]{revtex}

\begin{document}
\draft
\title{Random Bond Effect in the Quantum Spin System (Tl$_{1-x}$K$_{x}$)CuCl$_3$}

\author{Akira Oosawa and Hidekazu Tanaka}

\address{Department of Physics, Tokyo Institute of Technology, Oh-okayama, Meguro-ku, Tokyo 152-8551}

\date{\today}

\maketitle

\begin{abstract}
The effect of exchange bond randomness on the ground state and the field-induced magnetic ordering was investigated through magnetization measurements in the spin-$\frac{1}{2}$ mixed quantum spin system (Tl$_{1-x}$K$_{x}$)CuCl$_3$ for $x<0.36$. Both parent compounds TlCuCl$_3$ and KCuCl$_3$ are coupled spin dimer systems, which have the singlet ground state with excitation gaps ${\Delta}/k_{\rm B}=7.7$ K and 31 K, respectively. Due to bond randomness, the singlet ground state turns into the magnetic state with finite susceptibility, nevertheless, the excitation gap remains. Field-induced magnetic ordering, which can be described by the Bose condensation of excited triplets, magnons, was observed as in the parent systems. The phase transition temperature is suppressed by the bond randomness. This behavior may be attributed to the localization effect.

\end{abstract}


\section{Introduction}

In the last two decades, extensive theoretical studies on the effects of the exchange bond randomness in the one-dimensional quantum spin system have been performed \cite{Dasgupta,Fisher,Hyman}, and new concepts such as the random singlet (RS) phase \cite{Dasgupta,Fisher} and random dimer (RD) phase \cite{Hyman} have been proposed. For the spin-$\frac{1}{2}$ dimerized chain, Hyman {\it et al.} \cite{Hyman} showed that for weak randomness the excitation gap $\Delta$ does not close up, and the susceptibility remains finite, while for strong randomness the system is in the RD phase, in which the excitation spectrum is gapless and the susceptibility diverges as $\chi \sim T^{-1+\alpha}$ with $0 < \alpha \ll 1$. By means of magnetic susceptibility and specific heat measurements, Manaka {\it et al.} \cite{Manaka} investigated the bond randomness effect in the $S=1/2$ alternating Heisenberg chain system (CH$_3$)$_2$CHNH$_3$Cu(Cl$_x$Br$_{1-x}$)$_3$. They observed that with increasing $x$, the ordered phase with high ordering temperature appears discontinuously for $0.44 < x < 0.87$. 

The random bond effect in the three-dimensional (3D) spin gap system has been less extensively studied. This paper is concerned with the random bond effect in the 3D spin-$\frac{1}{2}$ mixed system (Tl$_{1-x}$K$_{x}$)CuCl$_3$. The parent compounds TlCuCl$_3$ and KCuCl$_3$ have the same monoclinic structure (space group $P2_1/c$) \cite{Willett,Takatsu}. The crystal structure is composed of planar dimers of Cu$_2$Cl$_6$, in which Cu$^{2+}$ ions have spin-$\frac{1}{2}$. The dimers are stacked on top of one another to form infinite double chains parallel to the crystallographic $a$-axis. These double chains are located at the corners and center of the unit cell in the $b-c$ plane, and are separated by Tl$^+$ or K$^+$ ions. The lattice parameters of both parent systems are shown in Table \ref{table1}. The crystal lattice of TlCuCl$_3$ is compressed along the $a$-axis and enlarged in the $b-c$ plane as compared with KCuCl$_3$. Thus, substituting K$^+$ ions for some Tl$^+$ ions produces not only exchange randomness, but also negative uniaxial stress along the $a$-axis. 

The magnetic ground states of TlCuCl$_3$ and KCuCl$_3$ are the spin singlet \cite{Tanaka1,Takatsu} with excitation gaps $\Delta=7.7$ \cite{Shiramura,Oosawa1} and 31 K \cite{Shiramura}, respectively. From the results of analyses of the dispersion relations obtained by neutron inelastic scattering, it was found that the origin of the spin gaps in TlCuCl$_3$ and KCuCl$_3$ is the strong antiferromagnetic interaction in the chemical dimer Cu$_2$Cl$_6$, and that the neighboring dimers couple magnetically along the double chain and in the $(1, 0, -2)$ plane, in which the hole orbitals of Cu$^{2+}$ spread \cite{Kato1,Cavadini1,Kato2,Cavadini2,Cavadini3,Kato3,Cavadini4,Oosawa2,Oosawa3}. TlCuCl$_3$ and KCuCl$_3$ differ in their interdimer interactions.  In TlCuCl$_3$ the interdimer interactions are strong \cite{Cavadini4,Oosawa2,Oosawa3}, while they are weak in KCuCl$_3$ \cite{Kato1,Cavadini1,Kato2,Cavadini2,Cavadini3,Kato3}.

When a strong magnetic field higher than the gap field $H_{\rm g}=\Delta/g\mu_{\rm B}$ is applied in the 3D spin gap system, magnetic ordering occurs. It has been theoretically shown that the field-induced magnetic ordering in the 3D Heisenberg spin gap system can be described by the Bose-Einstein condensation (BEC) of excited triplets (magnons) \cite{Giamarchi,Nikuni,Wessel1,Wessel2}. Recently, field-induced magnetic ordering was observed in TlCuCl$_3$ \cite{Oosawa1,Oosawa4,Tanaka2,Oosawa5} and KCuCl$_3$ \cite{Oosawa5}. Nikuni {\it et al.} \cite{Nikuni} demonstrated that the temperature dependence of the magnetization and field dependence of the transition temperature observed in TlCuCl$_3$ can be described by the BEC of dilute magnons. The magnon BEC is a new concept of magnetic phase transition. Thus, it is of great interest to study how the field-induced magnetic ordering is affected by the bond randomness. For this purpose, we carried out magnetization measurements of (Tl$_{1-x}$K$_{x}$)CuCl$_3$. 

\section{Experimental Details}

We first prepared single crystals of TlCuCl$_3$ and KCuCl$_3$ by the Bridgman method. The details of their preparation have been reported in reference \cite{Oosawa1}. Mixing single crystals of TlCuCl$_3$ and KCuCl$_3$ in a ratio of $1-x:x$, we prepared (Tl$_{1-x}$K$_{x}$)CuCl$_3$ by the Bridgman method. The temperature at the center of the furnace was set at 580$^{\circ}$C, and the lowering rate was 3 mm$\cdot$h$^{-1}$. Single crystals of 1$\sim$5 cm$^3$ with $x=0.04$, 0.05, 0.08, 0.11, 0.14, 0.16, 0.20, 0.27 and 0.36 were obtained. The potassium concentration $x$ was determined by emission spectrochemical analysis after the measurements. Samples used for magnetic measurements were cut into pieces of 50$\sim$150 mg. We treated samples in a glove box filled with dry nitrogen to reduce the amount of hydrate phase, which produces paramagnetic susceptibility at low temperatures, on the sample surface.

The magnetizations were measured down to 1.8 K in magnetic fields up to 7 T using a SQUID magnetometer (Quantum Design MPMS XL). The magnetic fields were applied perpendicular to the cleavage planes (0,1,0) and (1,0,${\bar 2}$). 

\section{Results and Discussion}

\begin{figure}[ht]
\vspace*{5cm}
\begin{minipage}{7.5cm}
 \epsfxsize=75mm
  \centerline{\epsfbox{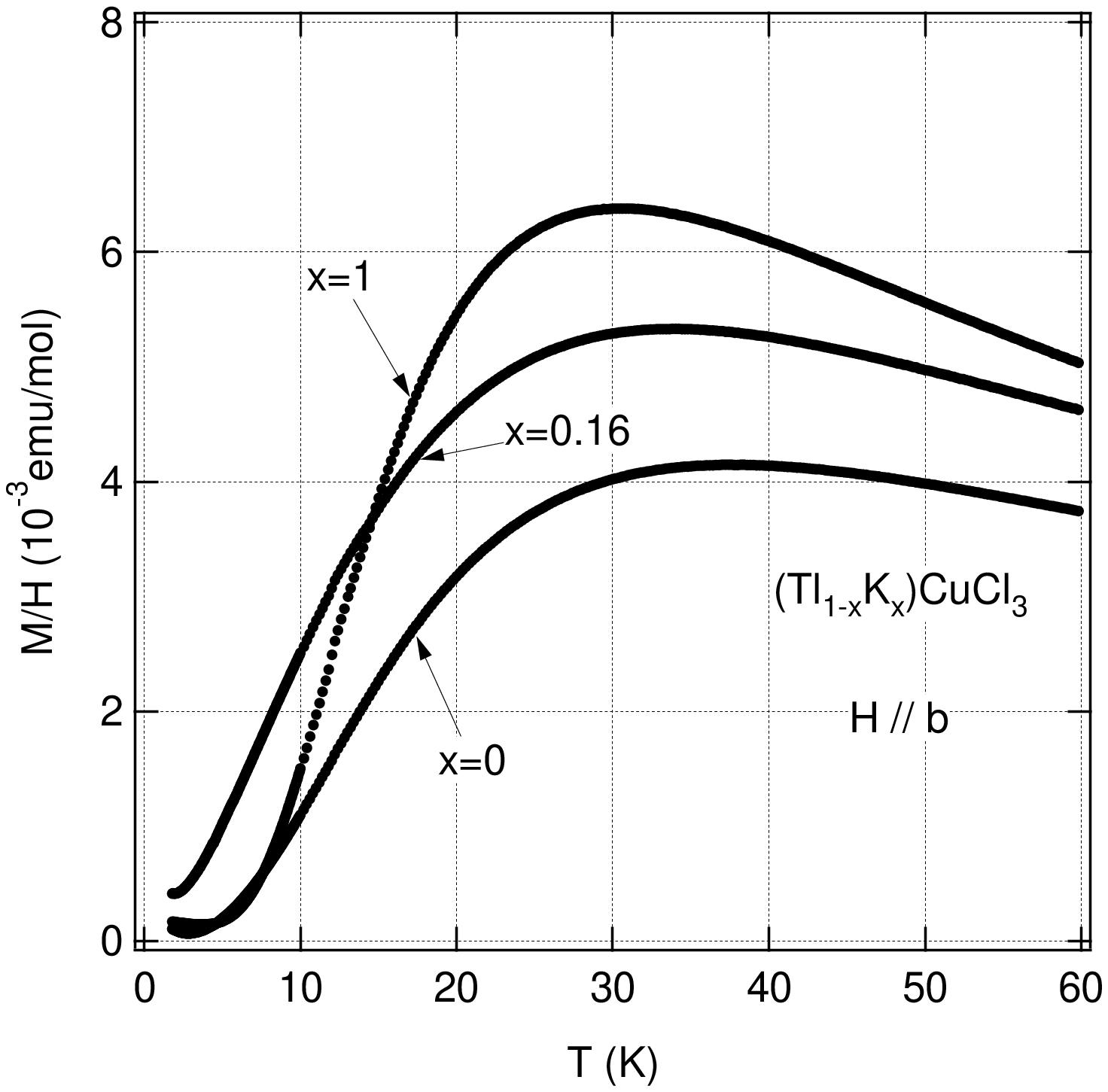}}
\begin{center}
(a)
\end{center}
\end{minipage}
\begin{minipage}{7.5cm}
 \epsfxsize=75mm
  \centerline{\epsfbox{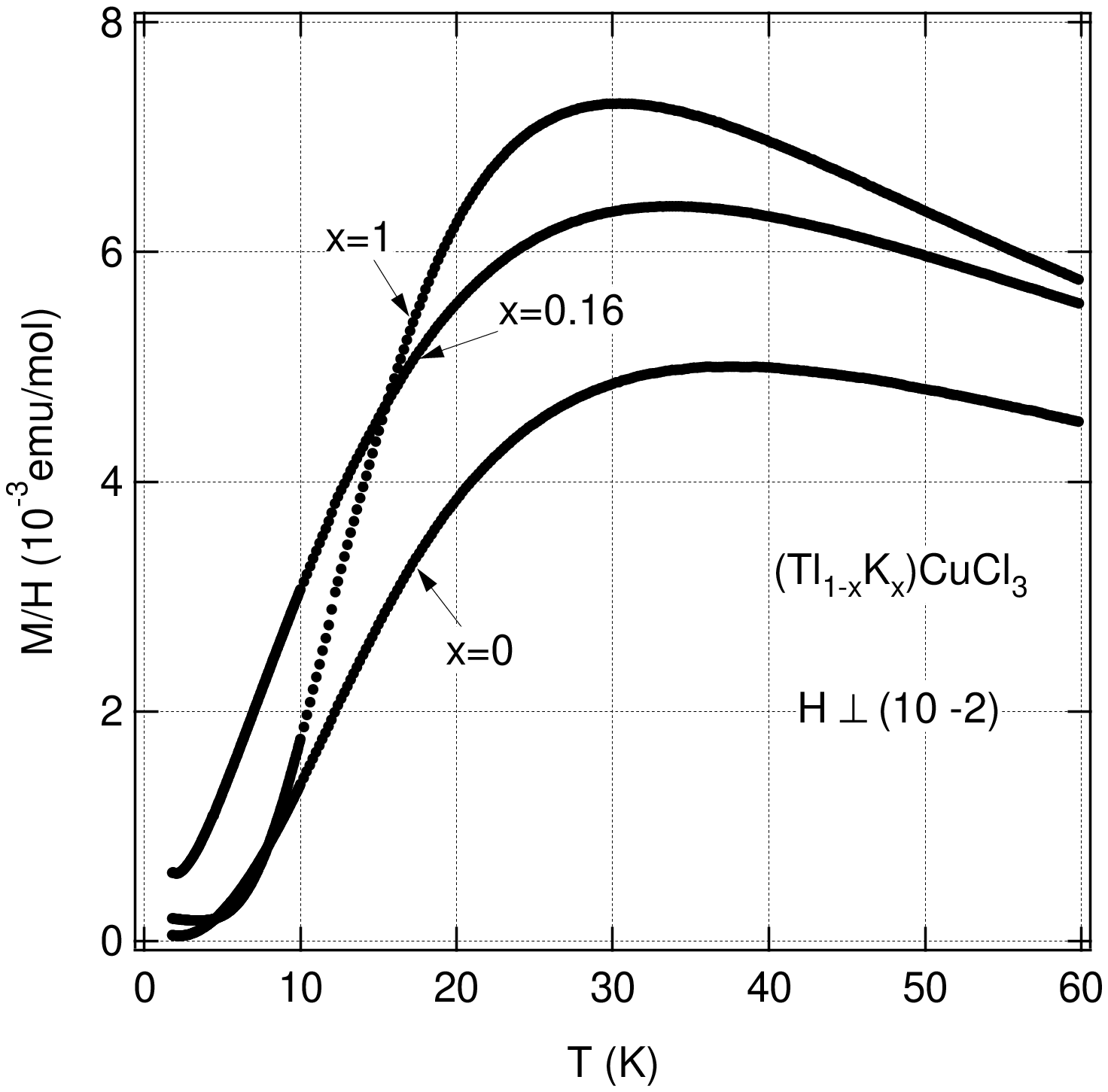}}
\begin{center}
(b)
\end{center}
\end{minipage}
\vspace*{1cm}
	\caption{The temperature dependence of the magnetic susceptibility $\chi=M/H$ in (Tl$_{1-x}$K$_{x}$)CuCl$_3$ for $x=0$, 0.16 and 1.00 for (a) $H\parallel b$ and (b) $H\perp (1,0,{\bar 2})$. The applied magnetic field is $H=1$ T for $x=0$ and 1, and $H=0.1$ T for $x=0.16$.}
	\label{Kfig1}
\end{figure}

Figure \ref{Kfig1} shows the temperature dependence of the magnetic susceptibilities ${\chi}=M/H$ in (Tl$_{1-x}$K$_{x}$)CuCl$_3$ for $x=0$ and 1 at $H=1$ T and for $x=0.16$ at $H=0.1$ T. The magnetic susceptibilities for pure systems ($x=0$ and 1) have broad maxima at $T=38$ K and $T=30$ K, respectively, and then decrease exponentially toward zero. This behavior is characteristic of a system with a gapped ground state. However, the magnetic susceptibility for $x=0.16$ decreases toward a finite value, obeying the power law, after it attains a broad maximum at $T=34$ K. 

\begin{figure}[ht]
\vspace*{5cm}
\begin{minipage}{7.5cm}
 \epsfxsize=75mm
  \centerline{\epsfbox{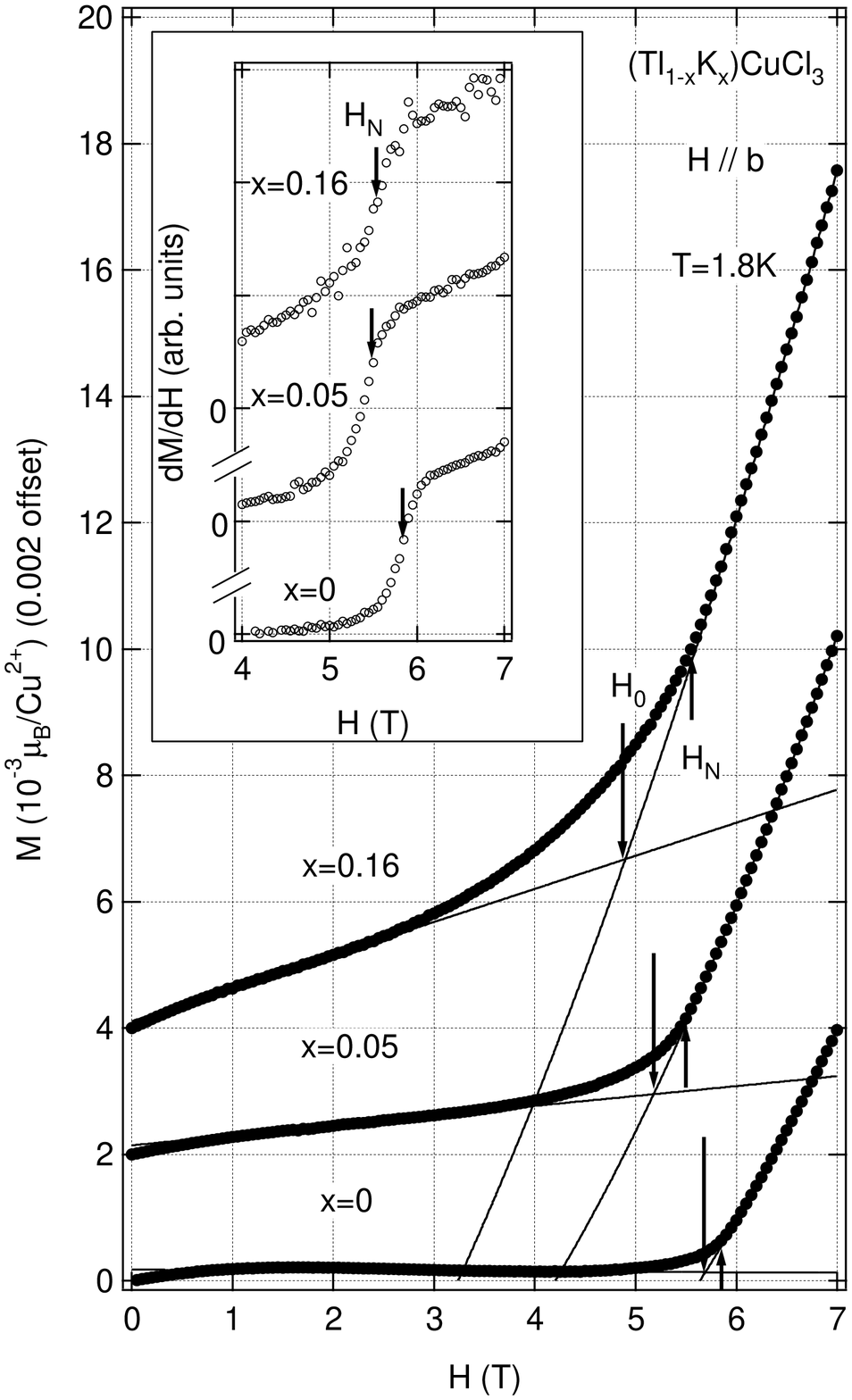}}
\begin{center}
(a)
\end{center}
\end{minipage}
\begin{minipage}{7.5cm}
 \epsfxsize=75mm
  \centerline{\epsfbox{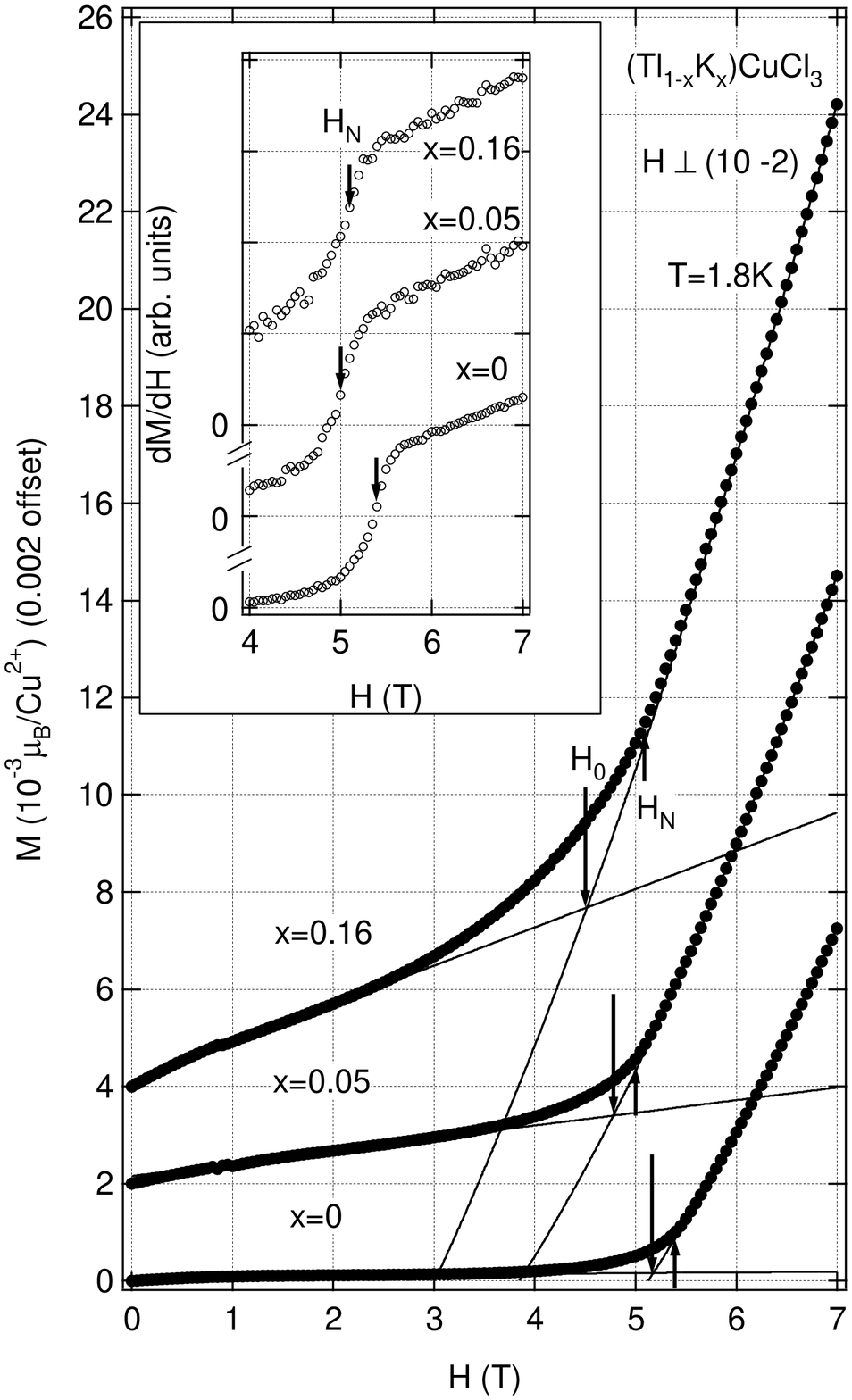}}
\begin{center}
(b)
\end{center}
\end{minipage}
\vspace*{1cm}
	\caption{The field dependence of the magnetization in (Tl$_{1-x}$K$_{x}$)CuCl$_3$ with $x=0$, 0.05 and 0.16 at $T=1.8$ K for (a) $H\parallel b$ and (b) $H\perp (1,0,{\bar 2})$. $H_{\rm N}$ denotes the phase transition field and $H_0$ is a field corresponding to the gap field $H_{\rm g}=\Delta/g\mu_{\rm B}$.}
	\label{Kfig2}
\end{figure}

Figure \ref{Kfig2} shows the magnetization curves in (Tl$_{1-x}$K$_{x}$)CuCl$_3$ measured at $T=1.8$ K for $x=0$, 0.05 and 0.16. In pure TlCuCl$_3$ ($x=0$), the magnetization is almost zero until the transition field $H_{\rm N}$, indicated by arrows, and then increases steeply. This magnetization behavior is typical of the spin gap system. However, for $x \neq 0$, the magnetization curves have a finite slope in the low-field region. This, together with the temperature variation of the susceptibility for $x \neq 0$ shown in Fig. 1, indicates that the ground state is not a spin singlet, but a magnetic with continuous excitations. The absence of the Curie term in the low-temperature susceptibility and the fact that the magnetization curves in the low-field region do not obey the Brillouin function, but are almost proportional to the applied field indicate that the ground state consists not of isolated or weakly coupled spins, but of strongly coupled spins. Therefore, the low-lying excitation may be spin-wave-like for $x \neq 0$. Although no anomaly indicative of magnetic ordering was observed down to 1.8 K at $H=0.1$ T, it may occur with a further decrease in temperature. 

With increasing magnetic field, the magnetization for $x \neq 0$ increases rapidly above $H_{\rm N}$. This indicates that field-induced magnetic ordering occurs for $H>H_{\rm N}$ as observed in a pure system. Thus, we can deduce that a gap remains in the excitation spectrum. Here, we assigned the fields with an inflection point of the magnetization field derivative as the transition field $H_{\rm N}$, as shown in the inset of Fig. 2. These observations that the energy gap does not close, and the susceptibility remains finite correspond to the argument by Hyman {\it et al.} \cite{Hyman} for weak bond randomness in the $S=1/2$ bond-alternating system, though the present system is a 3D system.

Since the paramagnetic susceptibility for $H < H_{\rm N}$ is very low, the energy gap $\Delta$ between the ground state and the first excited triplet may be expressed as $\Delta = g\mu_{\rm B}H_0$, where $H_0$ is a field at which the two fitting functions for the magnetizations for $H < H_{\rm N}$ and $H > H_{\rm N}$ cross, as shown in Fig. 2. For the fitting, we used a linear function for $H < H_{\rm N}$ and a quadratic function for $H > H_{\rm N}$, because $dM/dH$ for $H > H_{\rm N}$ is described by a linear function, as shown in the inset of Fig. 2. We assign $H_0$ obtained at $T=1.8$ K to the gap field $H_{\rm g}=\Delta/g\mu_{\rm B}$. The value of $(g/2)H_0\approx 5.8$ T obtained for TlCuCl$_3$ is consistent with the $(g/2)H_{\rm g}\approx 5.7$ T obtained from the specific heat and neutron scattering measurements \cite{Oosawa4,Tanaka2}. Since the magnetization curves at 1.8 K exhibit rounding around $H_{\rm N}$, there is a certain amount of error in the determination of $H_0$. 

\begin{figure}[ht]
\begin{minipage}{7.5cm}
 \epsfxsize=60mm
  \centerline{\epsfbox{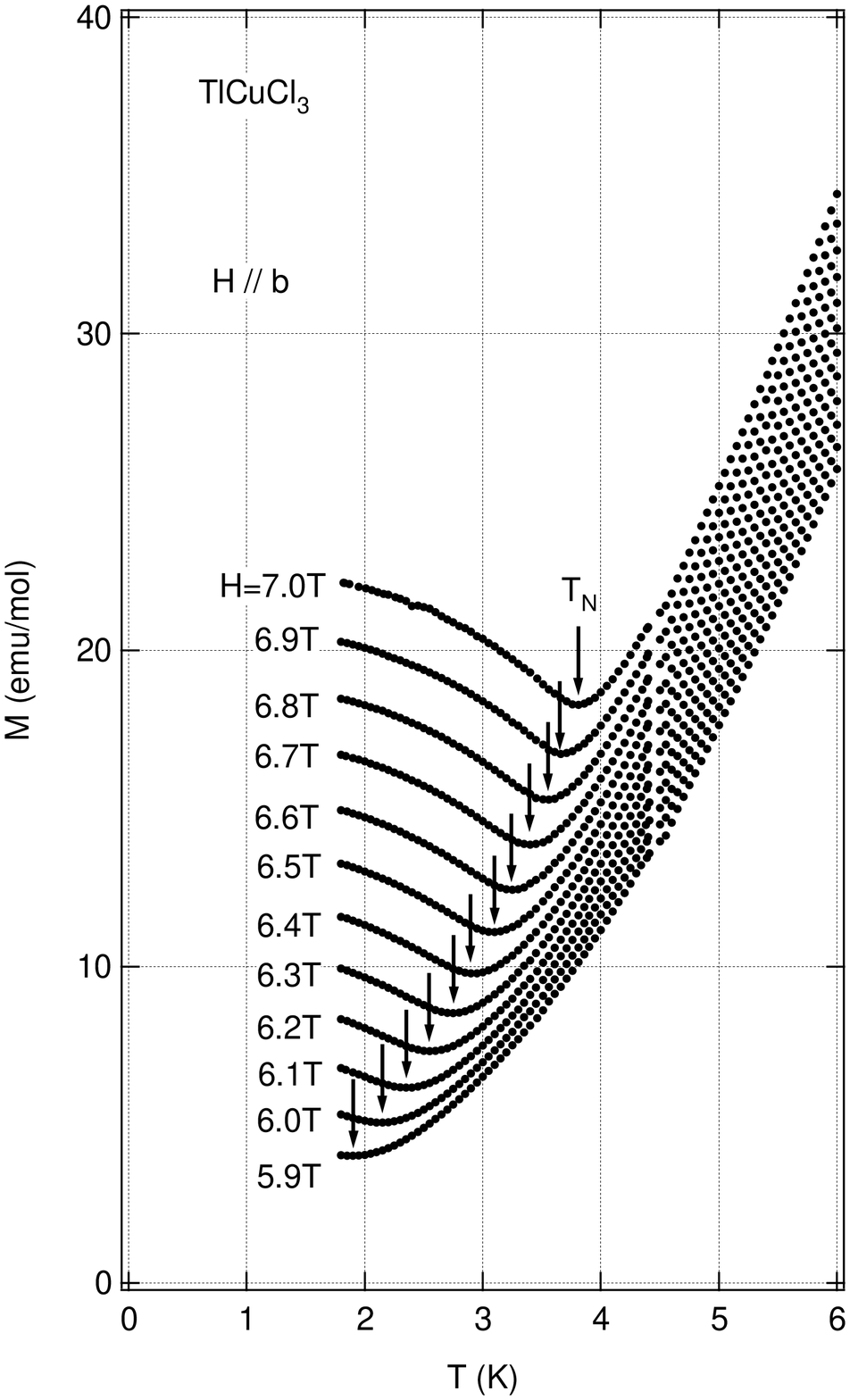}}
\begin{center}
(a)
\end{center}
\end{minipage}
\begin{minipage}{7.5cm}
 \epsfxsize=60mm
  \centerline{\epsfbox{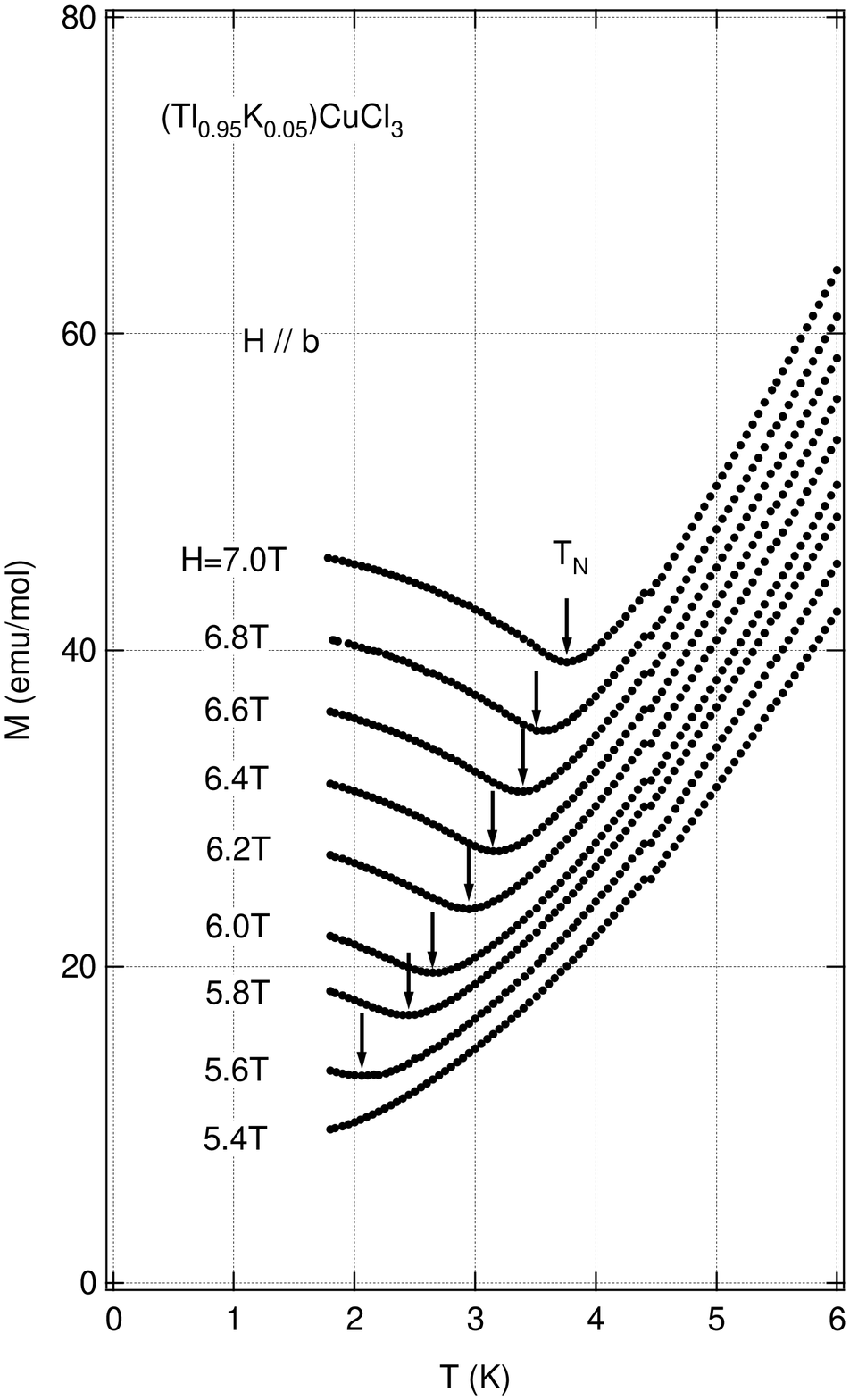}}
\begin{center}
(b)
\end{center}
\end{minipage}
\begin{minipage}{15cm}
 \epsfxsize=60mm
  \centerline{\epsfbox{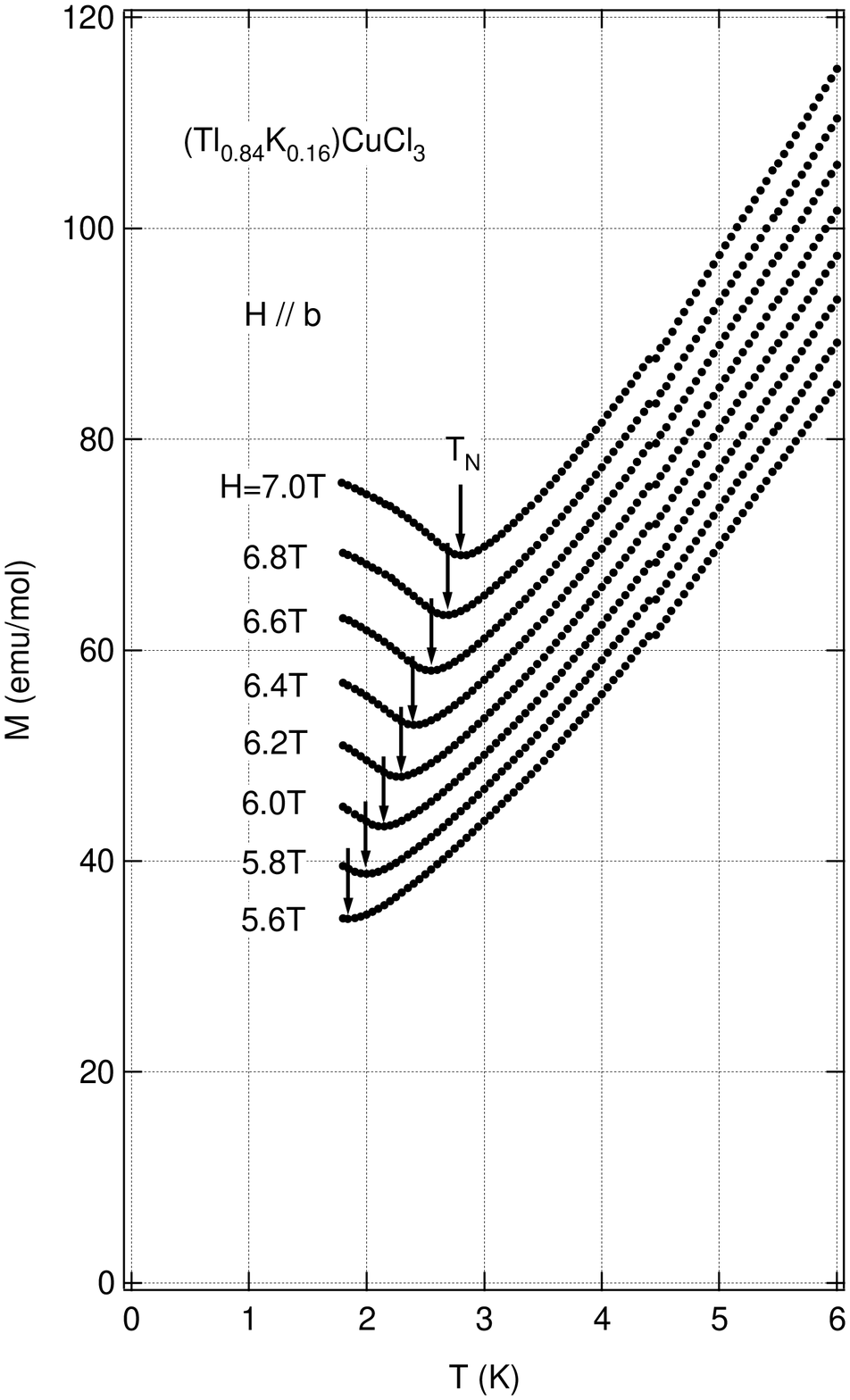}}
\begin{center}
(c)
\end{center}
\end{minipage}
\vspace*{1cm}
	\caption{The low-temperature magnetization in TlCuCl$_3$ and (Tl$_{1-x}$K$_{x}$)CuCl$_3$ for $x=0.05$ and 0.16 at various magnetic fields for $H\parallel b$.}
	\label{Kfig3}
\end{figure}

Figure \ref{Kfig3} shows the low-temperature magnetizations in (Tl$_{1-x}$K$_{x}$)CuCl$_3$ for $x=0.05$ and 0.16 at various magnetic fields. For comparison, we also show the low-temperature magnetizations for TlCuCl$_3$. The magnetic field was applied along the $b$-axis. The small anomalies at 4.4 K are due to an instrumental problem and are not intrinsic to the sample. TlCuCl$_3$ undergoes phase transition in magnetic fields higher than $H_{\rm g}\sim 6$ T, which is indicated by the cusplike minimum of the low-temperature magnetization \cite{Oosawa1,Oosawa4,Tanaka2}. The field-induced phase transition was also observed in KCuCl$_3$ for $H>H_{\rm g}\sim 22$ T \cite{Oosawa5}. 

The magnetization exhibits a similar cusplike minimum in (Tl$_{1-x}$K$_{x}$)CuCl$_3$ for $x=0.05$ and 0.16. The cusplike magnetization minimum in the present mixed systems is fairly sharp, similar to the pure system. This is indicative of the good homogeneity of the samples. We assign the temperature with the minimum magnetization as the transition temperature $T_{\rm N}$, which increases with increasing magnetic field. The transition temperature for $x=0.16$ is lower than that for $x=0.05$ at the same magnetic field. The magnetization minimum was not observed down to 1.8 K for $x=0.36$ at $H=7$ T. The low-temperature magnetization for $H\perp (1,0,{\bar 2})$ is similar to that for $H\parallel b$.
 
\begin{figure}[ht]
\vspace*{5cm}
\begin{minipage}{7.5cm}
 \epsfxsize=75mm
  \centerline{\epsfbox{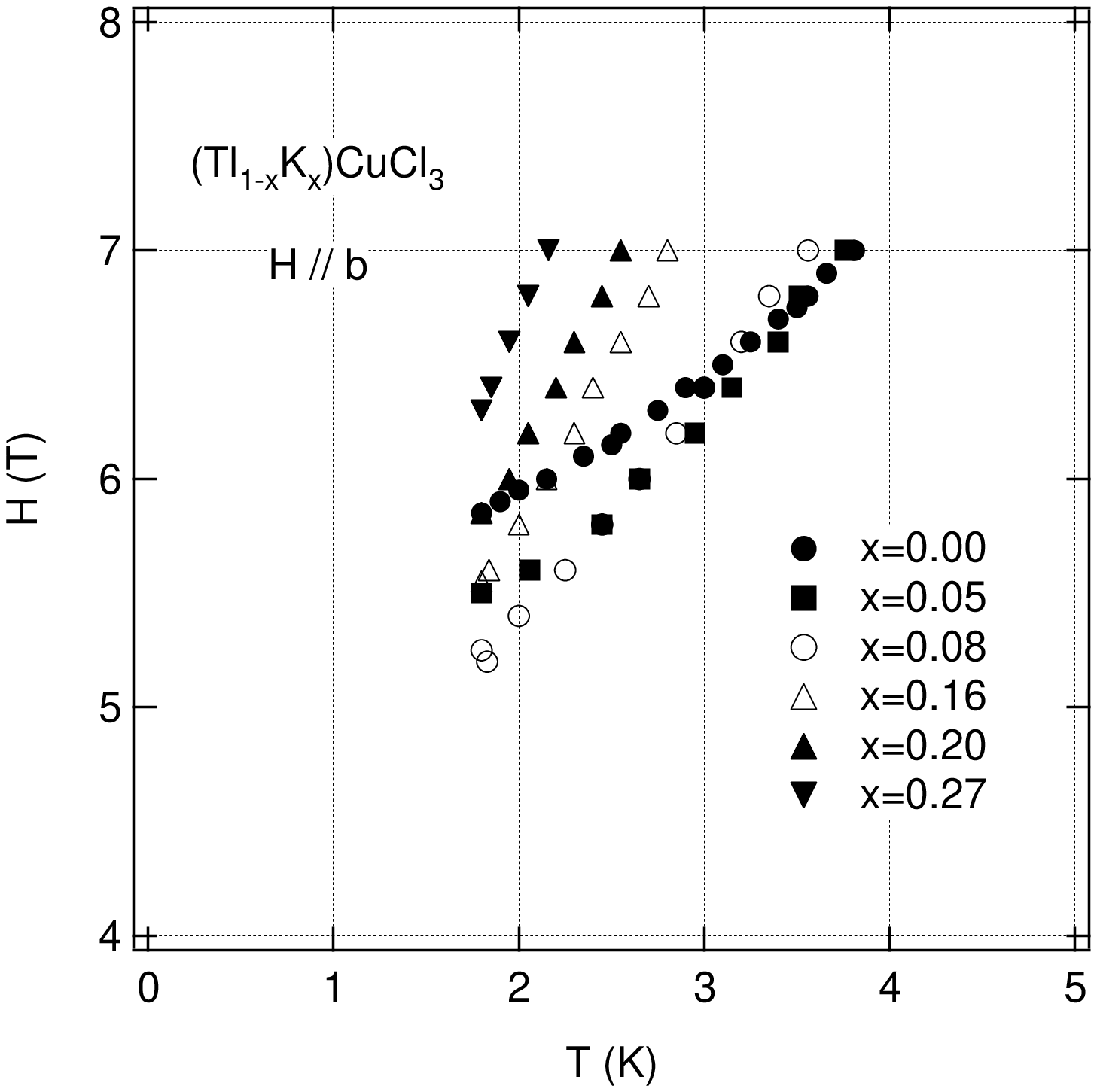}}
\begin{center}
(a)
\end{center}
\end{minipage}
\begin{minipage}{7.5cm}
 \epsfxsize=75mm
  \centerline{\epsfbox{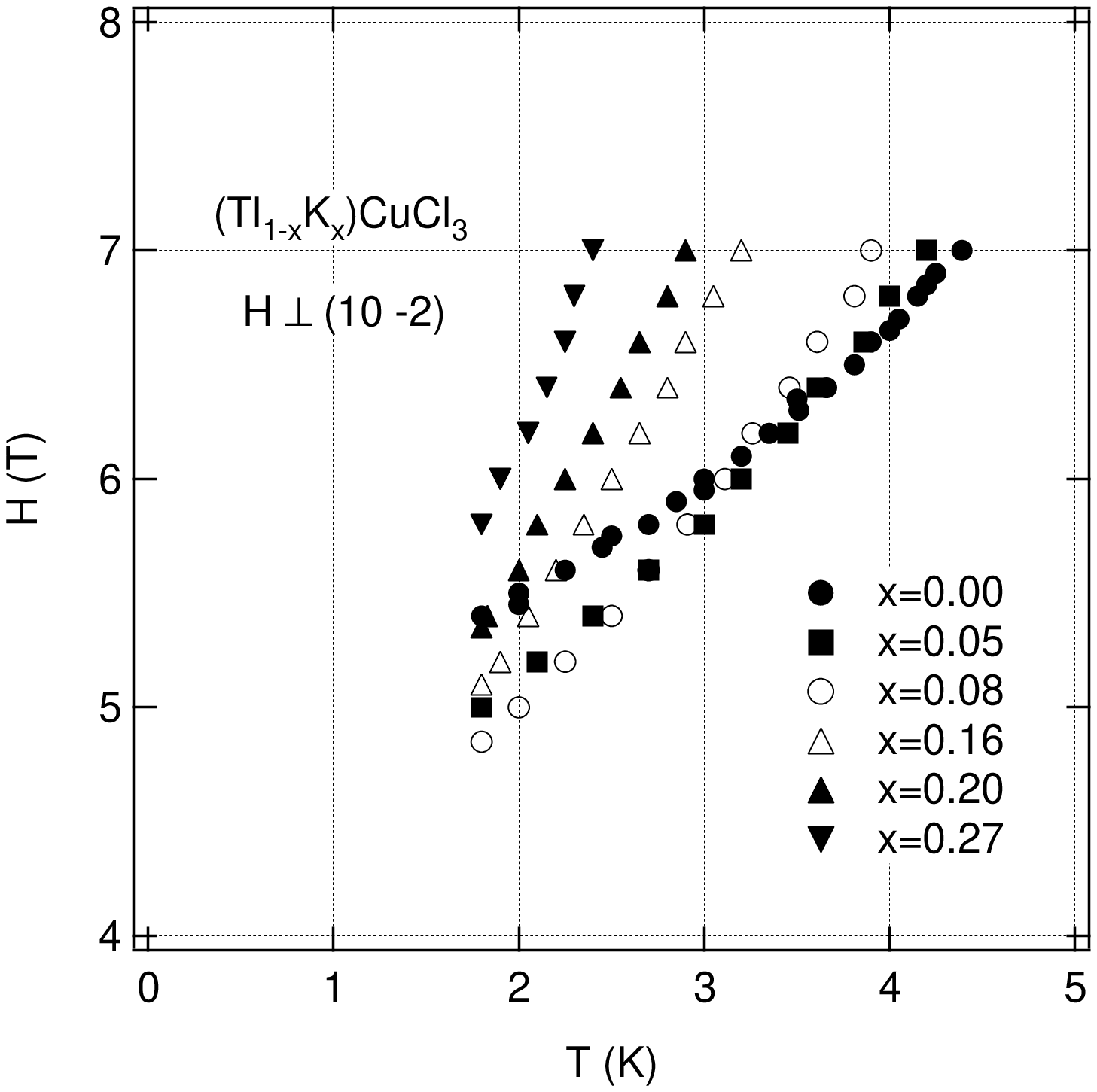}}
\begin{center}
(b)
\end{center}
\end{minipage}
\vspace*{1cm}
	\caption{The magnetic field vs temperature diagram for (Tl$_{1-x}$K$_{x}$)CuCl$_3$ with various concentrations $x$ for (a) $H\parallel b$ and (b) $H\perp (1,0,{\bar 2})$.}
	\label{Kfig4}
\end{figure}

Figure \ref{Kfig4} shows the magnetic field vs temperature diagram of (Tl$_{1-x}$K$_{x})$CuCl$_3$; the transition temperatures $T_{\rm N}$ and the transition fields $H_{\rm N}$ obtained for $H\parallel b$ and $H\perp (1,0,{\bar 2})$ are plotted for various potassium concentrations $x$. Since the $g$-factor is anisotropic, the values of $T_{\rm N}(H)$ and $H_{\rm N}(T)$ depend on the external field direction. Thus, we normalize the phase boundaries using $g=2.06$ for $H{\parallel}b$ and $g=2.23$ for $H{\perp}(1, 0, {\bar 2})$, which were obtained from TlCuCl$_3$ \cite{Oosawa1}.

\begin{figure}[ht]
\vspace*{5cm}
 \epsfxsize=90mm
  \centerline{\epsfbox{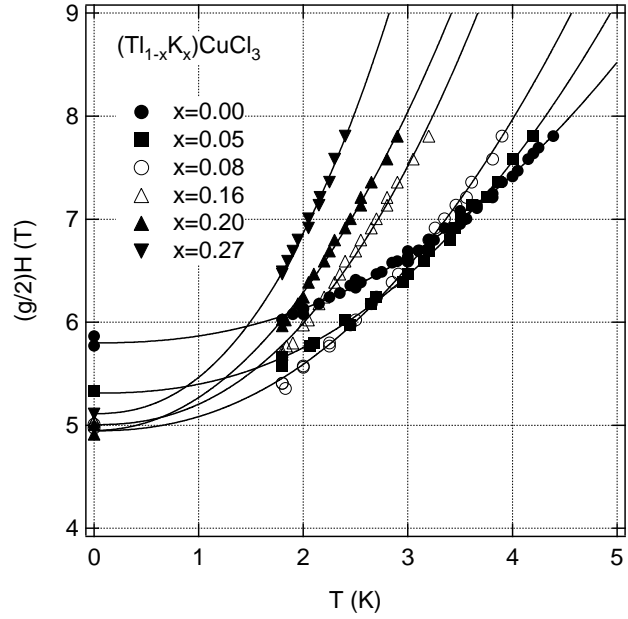}}
\vspace*{1cm}
	\caption{The phase diagram in (Tl$_{1-x}$K$_{x}$)CuCl$_3$ normalized by the $g$-factor. The solid lines denote the fitting with the power law.}
	\label{Kfig5}
\end{figure}

Figure \ref{Kfig5} shows the phase diagram of (Tl$_{1-x}$K$_{x})$CuCl$_3$ normalized by the $g$-factor. We also plot the value of the gap field $H_{\rm g}=\Delta/g\mu_{\rm B}$ at $T=0$ K, because 3D ordering is expected to occur at $T=0$ K when the energy gap vanishes. As seen from Fig. 5, the phase boundaries for $H{\parallel}b$ and $H{\perp}(1, 0, {\bar 2})$ coincide. This result indicates that the phase boundary is independent of the external field direction when normalized by the $g$-factor. This means that the magnetic anisotropy is negligible in (Tl$_{1-x}$K$_{x})$CuCl$_3$. The solid lines in Fig. \ref{Kfig5} denote the curves fitted using the power law
$(g/2)[H_{\rm N} (T) - H_{\rm N} (0)] \propto T^{\phi}$
which describes the phase boundary in TlCuCl$_3$ with $\phi=2.0\sim 2.2$ \cite{Nikuni,Oosawa4,Tanaka2}.
The values of the exponent $\phi$ and the normalized gap field $(g/2)H_{\rm N}(0)=(g/2)H_{\rm g}$ are plotted in Fig. 6 as a function of the concentration $x$. The exponent $\phi$ is almost independent of $x$, $\phi\approx 2.3$, while the magnitude of the normalized gap field $(g/2)H_{\rm g}$ decreases gradually to a constant $(g/2)H_{\rm g}\approx 5$ T. Since there is a certain amount of error in determining $H_{\rm g}$, as mentioned above, the value of $\phi$ has also a certain amount of error. When $x$ is close to unity, the normalized gap field $(g/2)H_{\rm g}$ may increase steeply and reach $(g/2)H_{\rm g}\approx 23$ T, which is the gap field for KCuCl$_3$ \cite{Shiramura}. 

Figure 7 shows the phase diagram of (Tl$_{1-x}$K$_{x})$CuCl$_3$, where the magnetic field is measured from the gap field $H_{\rm g}$. The transition temperature $T_{\rm N}$ decreases monotonically with increasing concentration $x$, namely, the range of ordered phase becomes narrower. In the pure system without randomness, the temperature and field ranges of the ordered phase are given by the interdimer interactions \cite{Tachiki1,Tachiki2}. The difference between the slope of the magnetization curve for $H > H_{\rm g}$ and that for the low-field region below $H_{\rm g}$ are given by the interdimer exchange interactions. In the pure system, the slope for the low-field region is negligible. The difference between the slopes is only nominally affected at potassium concentrations $x<0.3$, which implies that the average of the interdimer interactions does not decrease with $x$. Therefore, it is deduced that the decrease of the ordering temperature with $x$ is attributed to the exchange randomness due to the partial K$^+$ ion substitution. 

Nikuni {\it et al.} \cite{Nikuni} demonstrated that the field-induced phase transition in TlCuCl$_3$ can be described by the Bose-Einstein condensation (BEC) of excited triplets (magnons), {\it i.e.}, the theory represents the phase transition as the formation of the coherent state of magnons. The cusplike minimum of the magnetization, which corresponds to the magnon density, is typical of the BEC of dilute magnons \cite{Giamarchi,Nikuni,Wessel2}. Since the magnetization in the present mixed systems (Tl$_{1-x}$K$_{x})$CuCl$_3$ exhibits a similar cusplike minimum, the magnetic ordering may also be described by the magnon BEC. From the viewpoint of the magnon BEC, we discuss the decrease in the transition temperature with increasing potassium concentration $x$. 

Partial K$^+$ ion substitution for Tl$^+$ ions produces randomness in the exchange interactions between Cu$^{2+}$ ions. The magnitude of the randomness may increase with increasing $x$ for small $x$. The chemical potential $\mu$ of the magnon is expressed by $\mu=g\mu_{\rm B}(H-H_{\rm g})$ \cite{Nikuni}, where the gap energy $g\mu_{\rm B}H_{\rm g}$ is given by the intradimer and interdimer exchange interactions. The intradimer exchange interaction was evaluated, through neutron inelastic scattering experiments, to be $J\approx 5.7$ meV for TlCuCl$_3$ \cite{Cavadini4,Oosawa2,Oosawa3} and $J\approx 4.3$ meV for KCuCl$_3$ \cite{Cavadini1,Cavadini2,Kato3}. Since these two intradimer interactions are different, the partial K$^+$ ion substitution for Tl$^+$ ions produces the random on-site potential $\delta\mu_i$ of the magnon, where $i$ denotes the number of dimer sites. The random on-site potential acts to localize the magnon, {\it Anderson localization}. The hopping amplitude $t_{ij}$ and the intersite interaction $U_{ij}$ are given by the interdimer exchange interactions. Consequently, the partial K$^+$ ion substitution also produces randomness in $t_{ij}$ and $U_{ij}$, which may also act to localize the magnon. In general, the localization effect prevents the bosons from forming the coherent state, {\it i.e.}, the localization effect suppresses the BEC. Thus, we suggest that the localization effect due to the random bond acts to suppress the BEC of the magnons in the present system, such that the transition temperature $T_{\rm N}$ decreases with increasing $x$. However, we cannot have a quantitative argument on the decrease of the ordering temperature due to the randomness, because there is no quantitative theory on this problem, as far as we know. 

The superfluid-insulator transition in the boson system in the presence of the random on-site potential has been discussed theoretically \cite{Ma,Fisher2}. In our magnetic system, the superfluid phase corresponds to the ordered phase in which the magnons condense. In pure TlCuCl$_3$, the gapped phase for $H<H_{\rm g}$ corresponds to the insulating phase. Fisher {\it et al.} argued that a new phase, {\it the Bose glass phase}, can exist in the presence of randomness. The Bose glass phase is a phase in which bosons are localized, but there is no energy gap. It would be very interesting if a new phase corresponding to the Bose glass phase were observed. In order to confirm the presence of such a phase, more low-temperature measurements are needed.

\section{Conclusion}

By means of magnetization measurements, we have investigated the random bond effect in (Tl$_{1-x}$K$_x$)CuCl$_3$ which is a mixture of the three-dimensional coupled dimer systems TlCuCl$_3$ and KCuCl$_3$. Due to bond randomness, the singlet ground state turns into the magnetic state with finite susceptibility, but the excitation gap remains. For $x<0.3$, the magnitude of the gap field $H_{\rm g}$ decreases gradually to a constant. The field-induced magnetic ordering, which may be described by the Bose condensation of magnons, was observed, similar to TlCuCl$_3$. The ordering temperature is suppressed by bond randomness. We suggest that this behavior can be attributed to the localization effect due to randomness.
\acknowledgments
The authors would like to thank M. Oshikawa and T. Nikuni for useful discussions. This work was supported by Toray Science Foundation and a Grant-in-Aid for Scientific Research on Priority Areas (B) from the Ministry of Education, Culture, Sports, Science and Technology of Japan. A. O. was supported by the Research Fellowships of the Japan Society for the Promotion of Science for Young Scientists.

\begin{table}
\caption{Lattice constants $a, b, c$ and $\beta$ at room temperature for KCuCl$_3$ \cite{Willett} and TlCuCl$_3$ \cite{Tanaka2}.}
\label{table1}
\begin{tabular}{cccc} 
  & KCuCl$_3$ & & TlCuCl$_3$ \\ 
\tableline
$a$ $(\rm{\AA})$        & 4.029 & & 3.982 \\
$b$ $(\rm{\AA})$        & 13.785 & & 14.144 \\
$c$ $(\rm{\AA})$        & 8.736        & & 8.890 \\ 
$\beta$ & 97.33$^{\circ}$ & & 96.32$^{\circ}$ \\ 
\end{tabular}
\end{table}

\newpage

\newpage

\newpage

\newpage

\newpage

\newpage

\begin{figure}[ht]
\vspace*{5cm}
 \epsfxsize=90mm
  \centerline{\epsfbox{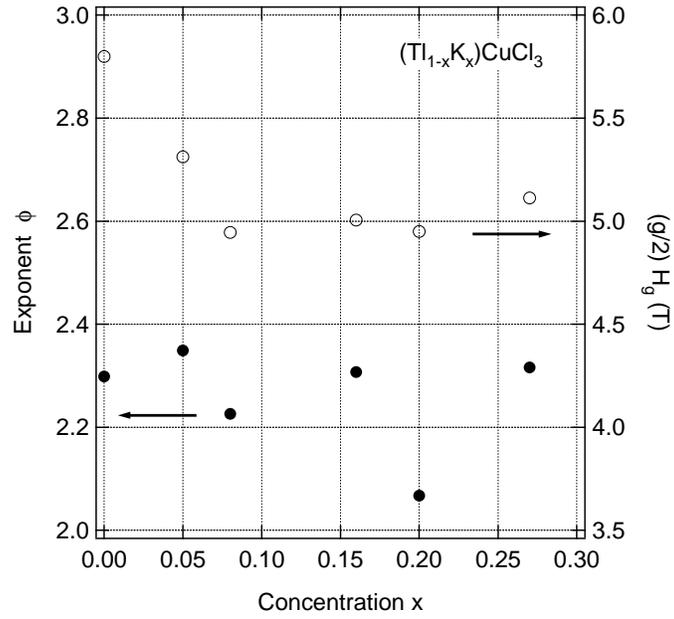}}
\vspace*{1cm}
	\caption{The exponent $\phi$ and the normalized gap field $(g/2)H_{\rm g}$ in (Tl$_{1-x}$K$_{x}$)CuCl$_3$ obtained by fitting the power law to the phase boundary.}
	\label{Kfig6}
\end{figure}      
	
\begin{figure}[ht]
\vspace*{5cm}
 \epsfxsize=90mm
  \centerline{\epsfbox{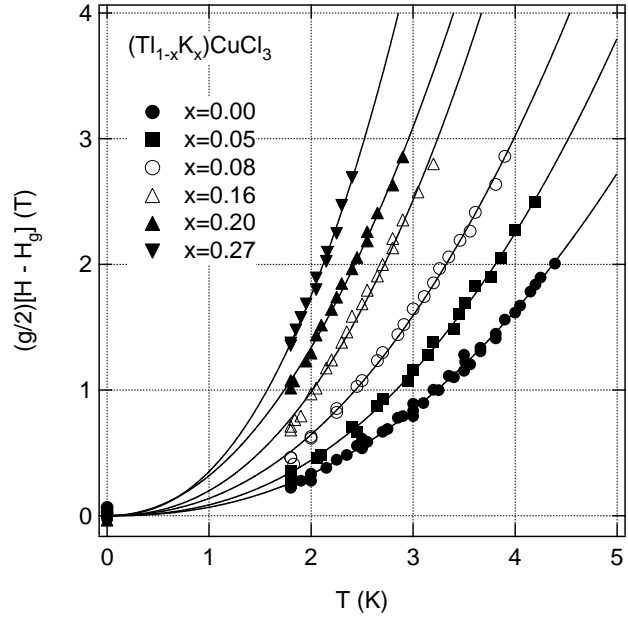}}
\vspace*{1cm}
	\caption{The phase diagram of (Tl$_{1-x}$K$_{x})$CuCl$_3$, where the magnetic field is measured from the gap field $H_{\rm g}$.}
	\label{Kfig7}
\end{figure}      

\end{document}